\documentclass[floatfix,twocolumn,showpacs,preprintnumbers,amsmath,amssymb,prl,superscriptaddress]{revtex4}
\newcommand{\etal}{{\it et al.}}
\usepackage{graphicx}
\usepackage{epstopdf}
\usepackage{dcolumn}

\begin{document}


\title{Electrical conductivity of high-pressure liquid hydrogen by quantum Monte Carlo methods}

\author{Fei Lin}
\affiliation{Department of Physics, University of Illinois at Urbana-Champaign, Urbana, IL 61801, USA}
\author{Miguel A. Morales}
\affiliation{Department of Physics, University of Illinois at Urbana-Champaign, Urbana, IL 61801, USA}
\author{Kris T. Delaney}
\affiliation{Materials Research Laboratory, University of California, Santa Barbara, CA 93106, USA}
\author{Carlo Pierleoni}
\affiliation{CNISM and Department of Physics, University of L'Aquila, L'Aquila, Italy}
\author{Richard M. Martin}
\affiliation{Department of Physics, University of Illinois at Urbana-Champaign, Urbana, IL 61801, USA}
\affiliation{Department of Applied Physics, Stanford University, Stanford, CA 94305, USA}
\author{D. M. Ceperley}
\affiliation{Department of Physics, University of Illinois at Urbana-Champaign, Urbana, IL 61801, USA}


\date{\today}
	
\begin{abstract}
We compute the electrical conductivity for liquid hydrogen at
high pressure using quantum Monte Carlo. The method uses
Coupled Electron-Ion Monte Carlo to generate configurations of
liquid hydrogen. For each configuration, correlated sampling of
electrons is performed in order to calculate a set of lowest
many-body eigenstates and current-current correlation functions
of the system, which are summed over in the many-body Kubo
formula to give AC electrical conductivity. The extrapolated DC
conductivity at 3000 K for several densities shows a liquid
semiconductor to liquid-metal transition at high pressure. Our
results are in good agreement with shock-wave data.
\end{abstract}


\maketitle

Liquid hydrogen at high pressure has been the subject of
intense experimental and theoretical research, because of its
special place in the periodic table and its cosmic abundance.
Understanding its behavior under high pressure and high
temperature is important for revealing the properties of giant
planets such as Jupiter and Saturn. Metallization of liquid
hydrogen at high pressure is of particular interest.
Experiments using shock waves have found a metallization
transition\cite{weir96}; at a pressure of 140 GPa and
temperature of 3000 K, liquid hydrogen has been reported to
turn from an insulating fluid into a metallic one with a DC
conductivity of about 2000 ($\Omega$ cm)$^{-1}$. However,
theoretically, such a metallization process has not been very
well understood. Mean-field density functional theory (DFT)
calculations \cite{hohl97} and Quantum Molecular Dynamics (QMD)
calculations \cite{collins01, redmer08} have been used to
calculate the electrical conductivity in liquid hydrogen, but
these methods neglect correlations between electrons. Such
correlations are likely to be important for the accurate
determination of transport properties. In this letter, we
propose and apply an alternative \emph{ab-initio} method which
combines the Coupled Electron-Ion Monte Carlo (CEIMC) method
\cite{pierleoni04} with the Correlation Function Quantum Monte
Carlo (CFQMC) method \cite{ceperley88, hammond94}.

Within the CEIMC approach \cite{pierleoni04, delaney06}, the
electrons and protons are simulated with separate but coupled
Monte Carlo simulations, taking advantage of the separation of
time scales in the Born-Oppenheimer approximation. The
Born-Oppenheimer energy surface for the protons is determined
using reptation quantum Monte Carlo (RQMC) \cite{baroni99}.
After the proton system reaches equilibrium, we record
uncorrelated samples of proton configurations, which are
subsequently used to determine the electrical conductivity.

The many-body Kubo formula \cite{kubo57} for
electrical conductivity is
\begin{eqnarray}
\sigma_{\alpha\alpha}(\omega)=& &\frac{2\pi e^2(1-e^{-\beta\hbar\omega})}{m^2\omega\Omega Z}
\sum_{k,n}|\langle k|\sum_i p_{\alpha}^i|n\rangle|^2\times \nonumber\\
& &e^{-\beta E_n}\delta(E_k-E_n-\hbar\omega),
\label{sigma}
\end{eqnarray}
where $\alpha=x,y,z$; $e$ $(m)$ is electron charge (mass),
$\Omega$ is the volume, $\beta$ is the inverse
temperature, $\omega$ is the frequency, and $p_{\alpha}^i$ is
the $\alpha$ component of the momentum operator for the $i$th
electron. $|n\rangle$ and $E_n$ are many-body eigen-states
and eigen-energies of the Hamiltonian, and $Z=\sum_n e^{-\beta E_n}$ the partition function.

The key challenge in evaluating the Kubo formula (Eq.
\ref{sigma}) is to determine the sum over all many-body
eigenstates of the system. We compute a number of the
lowest-energy eigenstates, and the relevant matrix elements,
using the CFQMC method \cite{ceperley88}, which we now explain.
Consider the subspace spanned by a set of $M$ many-body basis
states $f_j(\textbf{R})$, i.e.,
$\Phi_i(\textbf{R})=\sum_{j=1}^Md_{ij}f_j(\textbf{R}),$
where $\textbf{R}$ represents electronic coordinates. The
ground basis state, $f_0(\textbf{R})$, has Slater determinants
of Kohn-Sham single electron orbitals times a Jastrow pair
correlation and backflow corrections\cite{delaney06,
pierleoni08}. For the excited basis states $f_j(\textbf{R}),
j>0$, we use low-lying particle-hole excitations with respect
to the ground state $f_0(\textbf{R})$.

\begin{figure}[th]
\begin{center}
\includegraphics[clip,width=75mm]{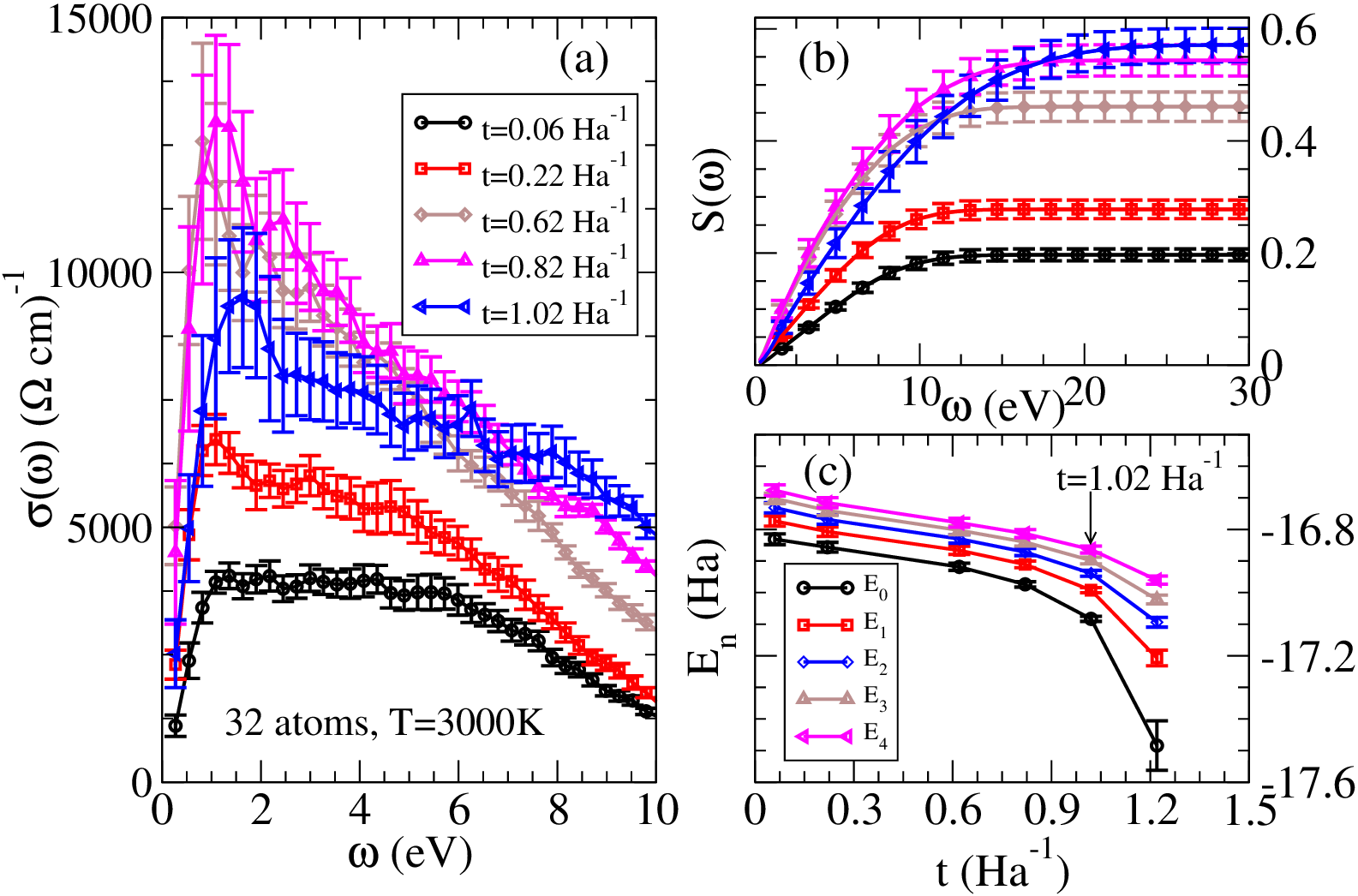}
\caption{(Color online). Electrical conductivity (a), the corresponding sum rule
(b), and five lowest energy levels (c) as a function of
imaginary-time projection $t$ for $r_s=1.40$. Note the decrease of $\sigma(\omega)$ at low
frequency is an artifact due to the gap in a finite system. The
simulation has $M=24$ excited states. The results are
averaged over 108 twist angles (see Fig. \ref{twistVSgamma}) and
10 decorrelated proton configurations drawn from the thermal Boltzmann
distribution sampled by our CEIMC calculations. Lines are just guides to the eye.} \label{sigmaVSbeta}
\end{center}
\end{figure}

\begin{figure}[th]
\begin{center}
\includegraphics[clip,width=75mm]{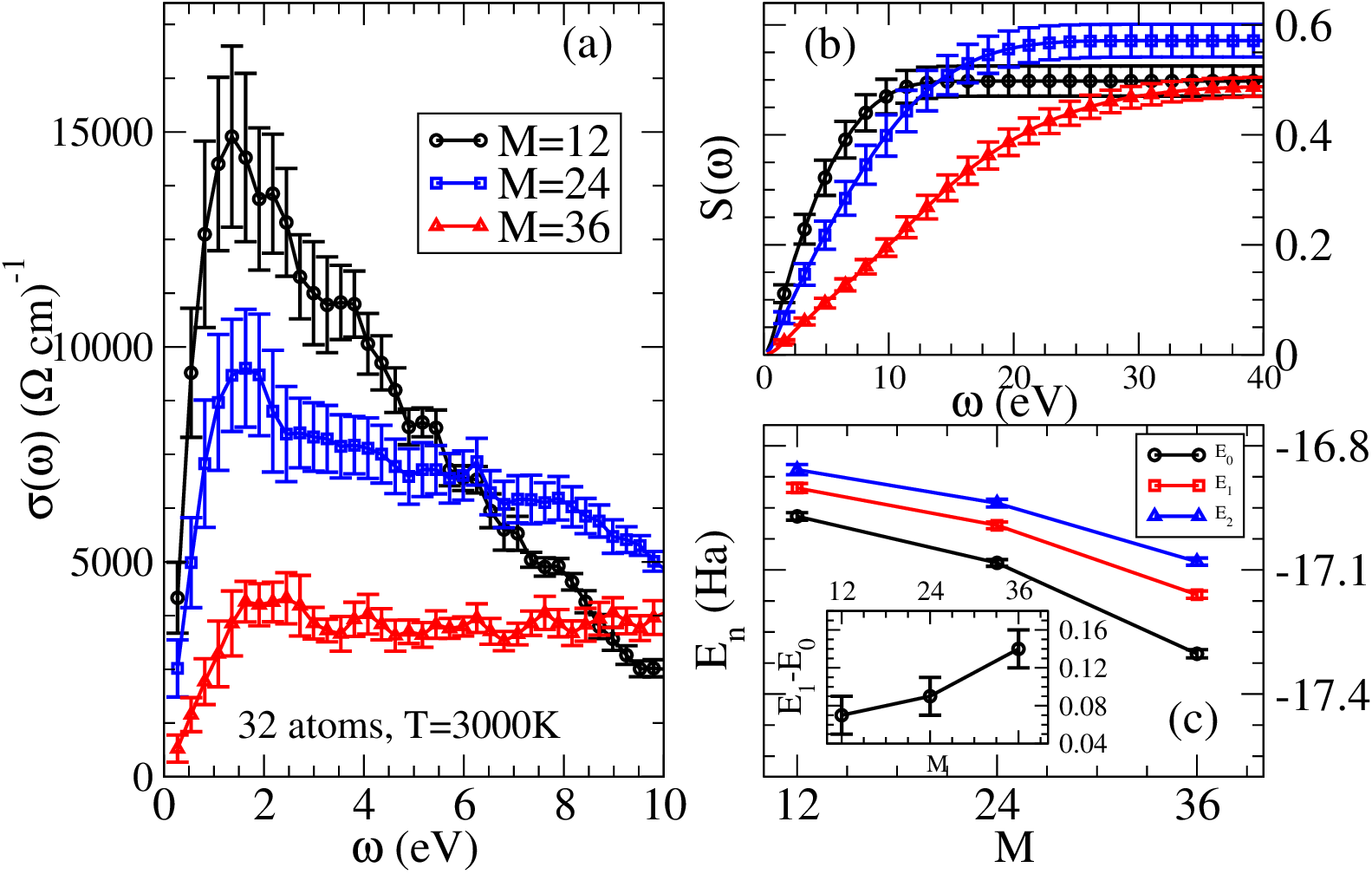}
\caption{(Color online). Electrical conductivity (a), the corresponding sum rule (b), and
first three lowest energy levels with energy difference $E_1-E_0$ in the inset as a function of
$M$, the number of trial wavefunctions in the basis set. Here $t=1.02$ Ha$^{-1}$ for $r_s=1.40$.
} \label{sigmaVSnexcited}
\end{center}
\end{figure}

Within this subspace, upper bounds to the exact eigenvalues can be found, by
minimizing the Rayleigh quotient with respect to $d_{ij}$,
\begin{equation}
\Lambda_i=\frac{\int d\textbf{R}\Phi_i^{*}(\textbf{R})\hat{H}\Phi_i(\textbf{R})}{\int d\textbf{R}\Phi_i^{*}(\textbf{R})\Phi_i(\textbf{R})},
\label{quotient}
\end{equation}
where $\hat{H}$ is the electronic Hamiltonian.
Furthermore, an improved basis set $\{\tilde{f}_i(\textbf{R})\}$ can be
obtained by applying an imaginary-time projection to the basis
set:
\begin{equation}
\tilde{f}_i(\textbf{R})=e^{-t\hat{H}/2}f_i(\textbf{R}),
\label{projectedbasis}
\end{equation}
where $t$ is the projection time.
Replacing $\{f_j(\textbf{R})\}$ with
$\{\tilde{f}_j(\textbf{R})\}$ and minimizing $\Lambda_i$ with
respect to $d_{ij}$, one obtains the many-body generalized eigenvalue equation:
\begin{equation}
\sum_{j=1}^M\textbf{H}_{ij}(t)d_{kj}(t)=\Lambda_k(t)\sum_{j=1}^M\textbf{S}_{ij}(t)d_{kj}(t),
\label{eigenequation}
\end{equation}
where $\Lambda_k(t)$ is the $k$th eigenvalue, and the Hamiltonian $\textbf{H}$ and the overlap
matrices $\textbf{S}$ are defined with respect to the projected basis set $\{\tilde{f}_i(\textbf{R})\}$ as
\begin{eqnarray}
\textbf{H}_{ij}&=&\int
d\textbf{R}_1d\textbf{R}_2f_j(\textbf{R}_2)\langle\textbf{R}_2|\hat{H}e^{-t\hat{H}}|\textbf{R}_1\rangle
f_i(\textbf{R}_1),\label{hmatrix}\\
\textbf{S}_{ij}&=&\int d\textbf{R}_1d\textbf{R}_2f_j(\textbf{R}_2)\langle \textbf{R}_2|e^{-t\hat{H}}|\textbf{R}_1\rangle f_i(\bf{R}_1).\label{smatrix}
\end{eqnarray}
Note that we have used Eq. (\ref{projectedbasis}) and inserted
complete sets of basis states $\{|\textbf{R}\rangle\}$. These
matrix elements are calculated all at once with RQMC.

Solving Eq. (\ref{eigenequation}) yields the many-body eigenvalues and eigenfunctions that are
required for the Kubo formula. The current-current correlation functions are
computed at half of the projection time $t/2$ in RQMC \cite{ceperley88}. Finally, the $\delta$ function
in the Kubo formula is broadened by a Gaussian function during the
numerical calculation, whose width is of the same
order as the observed many-body energy spacing. 

We use the ground state as guiding wavefunction during RQMC
simulations \footnote{It has been shown that an improved
guiding wavefunction exists in the Bosonic excited state
simulations \cite{ceperley88}. We tested this approach by
including excited states, which are assigned with some
appropriate weights, in the RQMC guiding wavefunction, but the
method becomes unstable, possibly due to the Fermion sign
problem.}. The method is able to determine the lowest-lying
states of the system, typically fewer than $50$ in our
simulations, after which the calculation becomes less efficient
due to the increased fluctuations in the matrix elements in
Eqs. (5-6), a problem very much related to the well-known
fermion sign problem of QMC. An analysis
shows\cite{ceperley88} that the Monte Carlo (MC) efficiency must decrease
with projection time as $\exp(-\alpha t)$, making convergence
in $t$ problematical:  we call this the ``efficiency problem''.
However, it is important to note that as the basis is increased
in size, the projected low energy states become more accurate
\cite{ceperley88}. This suggests that we should include as many
excitations as long as the efficiency is not severely reduced.

\begin{figure}[th]
\begin{center}
\includegraphics[clip,width=60mm]{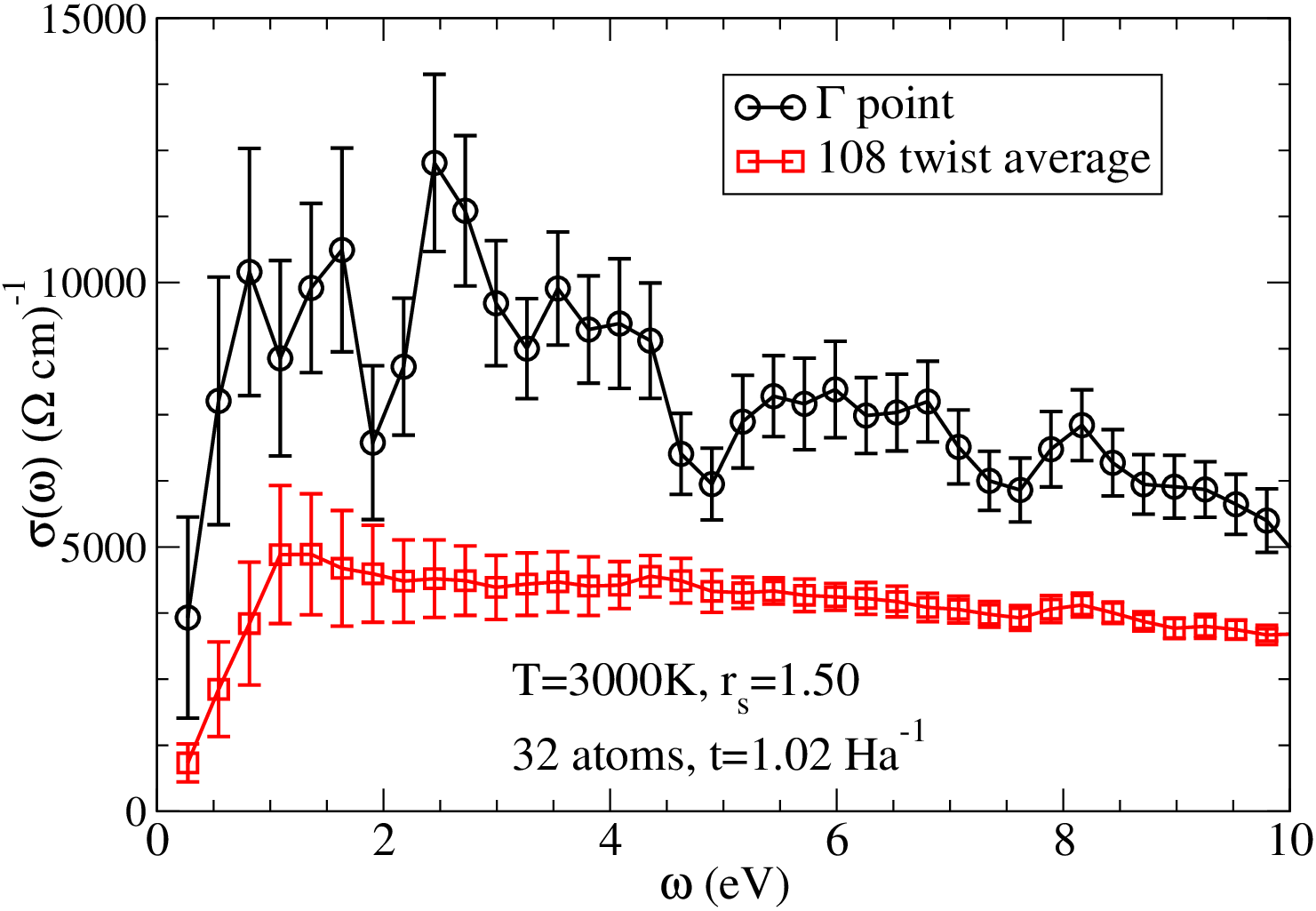}
\caption{(Color online). Electrical conductivity calculated at
the $\Gamma$ point compared with the twist-averaged one.}
\label{twistVSgamma}
\end{center}
\end{figure}

\begin{figure}[th]
\begin{center}
\begin{tabular}{c}
\includegraphics[clip,width=60mm]{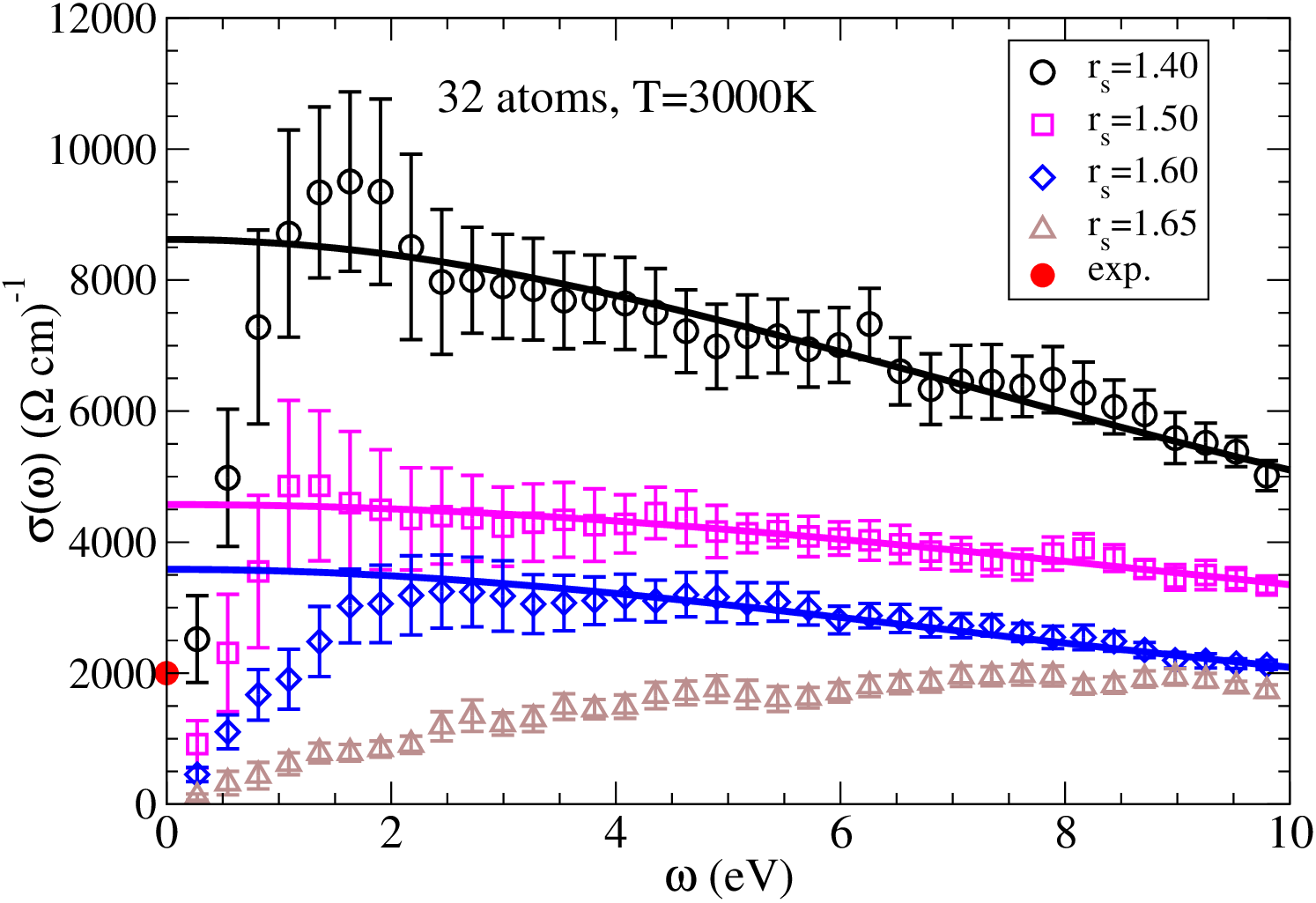} \\
\includegraphics[clip,width=60mm]{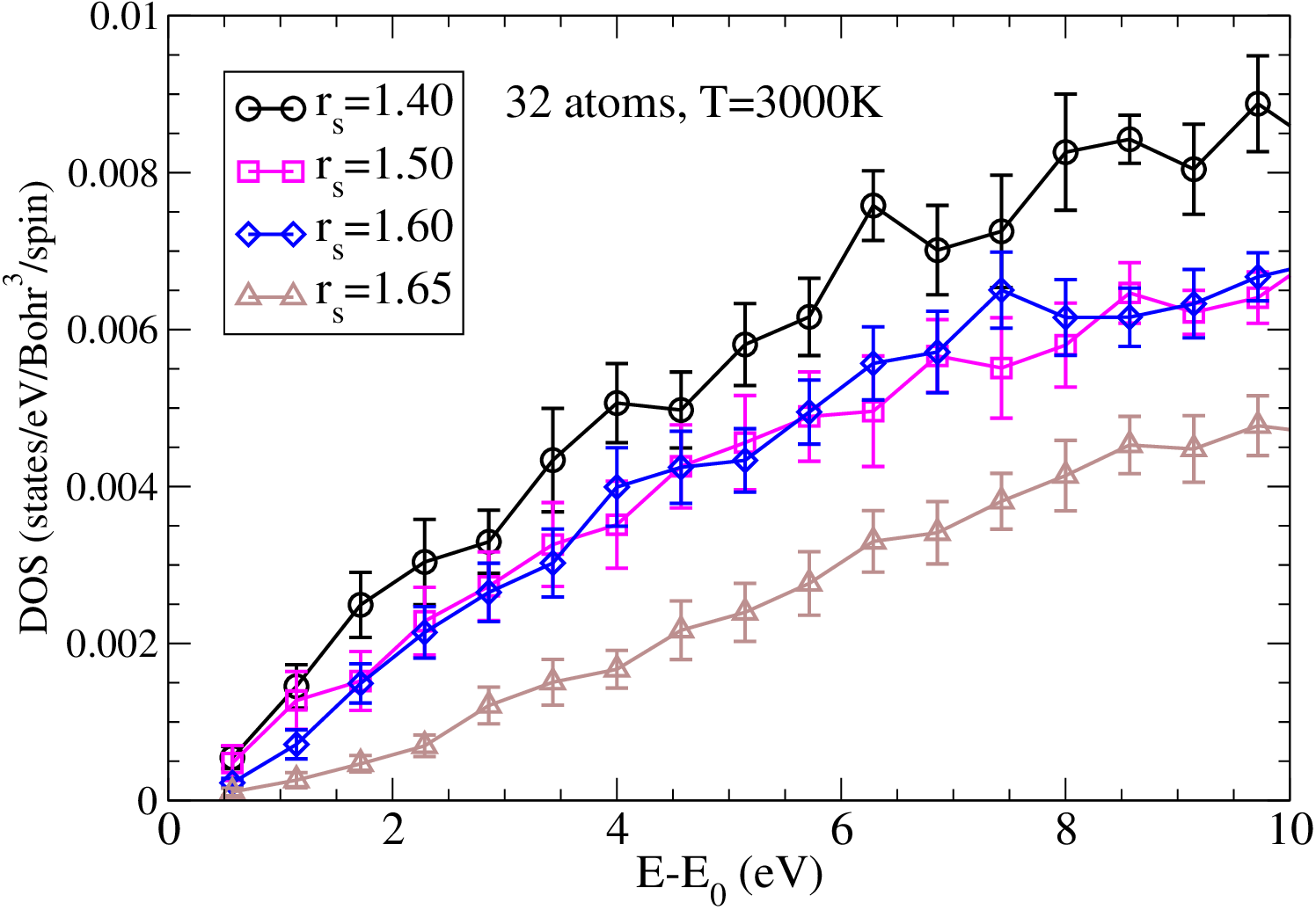}\\
\end{tabular}
\caption{(Color online). Electrical conductivity (scattered points) along
with the Drude fits (solid lines) (upper panel) and DOS (lower panel) as
a function of density for liquid hydrogen. The red dot at $\omega=0$ indicates the experimental DC
conductivity value after  the liquid hydrogen is metallized by
high pressure at 3000 K \cite{weir96}. The drop of electrical
conductivity to zero at zero frequency for $r_s=1.40$, $1.50$
and $1.60$ is due to finite-size effects.}
\label{sigmaandDOSat3K}
\end{center}
\end{figure}

All simulations reported here were at 3000 K, the temperature
of the shock experiments measuring conductivity \cite{weir96}. We first
investigate the role of RQMC imaginary-time projection $t$.
Fig. \ref{sigmaVSbeta} (a) shows the electrical conductivity of
liquid hydrogen  as a function of $t$. Defining
$S(\omega)=\frac{2m}{\pi
en_e}\int_0^{\omega}\sigma(\omega^{'})d\omega^{'},$
the conductivity sum rule is
$\lim_{\omega\rightarrow\infty} S(\omega)=1$ \cite{kubo57}. Here $n_e$ is
electron density. In the present method, we cannot include all
many-body states, so we do not expect the sum rule to be
satisfied. However, it provides an indication of calculation
quality. The main measure of appropriate projection time is the
convergence of low-frequency $\sigma(\omega)$. We can see from
Fig. \ref{sigmaVSbeta} (a) that low-frequency $\sigma(\omega)$
curves converge between the interval $t\in [0.82,
1.02]$Ha$^{-1}$. Here we choose the upper bound of $t=1.02 {\rm
Ha}^{-1}$, since we want to have as large $t$ as possible
before the efficiency is reduced to an intolerable level. Note
that the decrease in $\sigma(\omega)$ at $\omega < 2$ eV is an
artifact due to the finite number of atoms in the simulation, as we
discuss below. Fig. \ref{sigmaVSbeta} (b) also shows that the
best sum rule satisfaction is achieved at $t=1.02 {\rm
Ha}^{-1}$, where $S(\omega)\sim 0.6$ as
$\omega\rightarrow\infty$. As a further test of the $t$ value,
we show in Fig. \ref{sigmaVSbeta} (c) five lowest energy states
as a function of projection time. We see that indeed after
$t=1.02 {\rm Ha}^{-1}$, the statistical variances increase
substantially, and the estimates of the lowest energies are
pushed down due to the noise \cite{hammond94}.

Fig. \ref{sigmaVSnexcited} shows (a) the electrical
conductivity, (b) the corresponding $S(\omega)$, and (c) the
first 3 energy levels at fixed projection time as a function of
$M$, the size of the basis set. As before, we choose $M$ as
large as possible while still obtaining a reasonable MC
efficiency. To check the efficiency, we collect the same
amounts of stochastic data for the Hamiltonian and overlap
matrices in Eq. \ref{hmatrix} and \ref{smatrix}, solve the
generalized eigenvalue equation Eq. \ref{eigenequation}, and
look at the energy gap between ground state $E_0$ and the first
excited state $E_1$. A sudden increase of the energy gap
indicates the reduction of the efficiency, since noise can push
down the most the lowest energy level \cite{hammond94}. The
inset of Fig. \ref{sigmaVSnexcited} (c) shows an increased
slope for $E_1-E_0$ curve at $M=36$, indicating such a
reduction of efficiency. The corresponding $\sigma(\omega)$
curve is also suppressed in the low frequency region. See Fig.
\ref{sigmaVSnexcited} (a). 
We find that $M=24$ is a good compromise between accuracy and
efficiency. We also notice from Fig. \ref{sigmaVSnexcited} (b)
that $S(\omega)\rightarrow 0.6$ as $\omega\rightarrow\infty$
for $M=24$; while for $M=12$ and $36$, the corresponding sum
approaches $0.5$.

We note that our simulations 
are subject to finite-size errors in evaluating thermodynamic
properties. Such effects can be reduced by using twist-averaged
boundary conditions \cite{lin01}. 
In Fig. \ref{twistVSgamma} we compare $\sigma(\omega)$
calculated at the $\Gamma$ point of the Brillouin zone (BZ) to
the one averaged over a $6\times 6\times 6$ grid in the twist
angle space (the BZ).  We find that $\Gamma$ point sampling
overestimates the conductivity value.
However, finite-size effects have not been entirely eliminated
and are responsible for the observed vanishing of
$\sigma(\omega)$ at small $\omega$ for all systems studied.
This will not happen for a sufficiently large system in the
metallic phase (e.g. at $r_s\leq 1.50$). A common
procedure\cite{galli90, collins01, galli00} to estimate the DC
conductivity is to discard a few data points and extrapolate
the higher frequency data to $\omega=0$ . For systems inside
the metallic phase one can use the Drude formula
$\sigma(\omega)=\sigma_{\rm DC}/(1+\omega^2\tau^2)$
to estimate the DC conductivity $\sigma_{\rm DC}$ and the electronic relaxation
time $\tau$.

\begin{figure}[th]
\begin{center}
\begin{tabular}{c}
\includegraphics[clip,width=60mm]{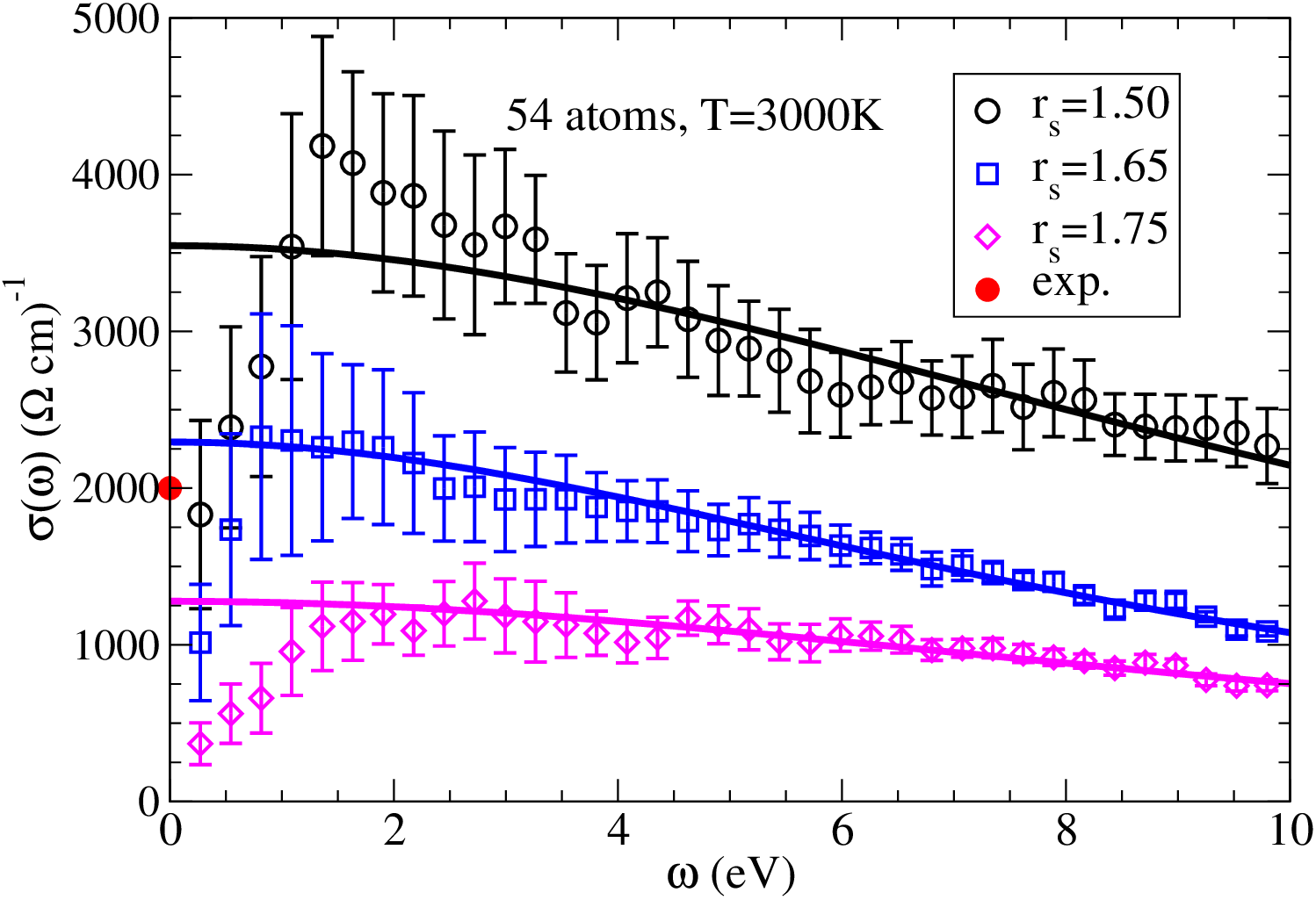}\\
\end{tabular}
\caption{(Color online). Electrical conductivity (scattered points) and Drude fits (solid lines) as a function
of density for liquid hydrogen. The red dot at $\omega=0$
denotes the experimental DC conductivity value after the liquid
hydrogen is metallized by high pressure \cite{weir96}.}
\label{sigmaandDOSat3KNP54}
\end{center}
\end{figure}
We now proceed to study the electrical conductivity at various
densities at T= 3000 K, where the metallization of liquid
hydrogen has been observed \cite{weir96}. We run CEIMC
simulations at $r_s=1.40$, 1.50, 1.60, and 1.65 for 32 protons
and electrons. Results (scattered points) along with the
corresponding Drude fits (solid lines) are shown on the upper
panel of Fig. \ref{sigmaandDOSat3K} and many-body density of
states (DOS) on the lower panel, where all the energy values
are calculated with respect to the ground-state energy of the
configuration. We can see that DC conductivity decreases from
about 8600 $(\Omega$ cm$)^{-1}$ (see below) at $r_s=1.40$ to
about $3500$ $(\Omega$ cm$)^{-1}$ at $r_s=1.60$ and to nearly
zero at $r_s=1.65$. The dot at $\omega=0$ in Fig.
\ref{sigmaandDOSat3K} shows the experimental DC conductivity
after metallization \cite{weir96}. The conductivity points for
$r_s=1.65$ near zero frequency has typical features for liquid
semiconductors, while a $2000$ $(\Omega$ cm$)^{-1}$ is a
typical liquid metal value \cite{shimoji77}. Thus, based on
32-atom simulations we see a liquid metal to a liquid
semiconductor transition in a density region of $1.60<r_s<1.65$
at T=3000 K for liquid hydrogen under high pressure, which is
close to the experimental transition density of $r_s=1.62$.
Such a metal-semiconductor transition is also hinted in the DOS
figure. Going from $r_s=1.40$ to $r_s=1.65$, one can see the
gradual opening of an energy gap.

We further look at the metallic behavior of the liquid hydrogen at $r_s=1.40$.
The Drude fit gives $\sigma_{\rm DC}=8620\pm 1000$ $(\Omega$
cm$)^{-1}$ and $\tau=2.0\pm 0.2$ a.u. or $3.1\pm 0.3\times
10^{-16}$ s. A rough estimate of electron mean free path ($l$)
then gives $l\sim 1.2$ \AA, which is about 1.5 times the
average proton-proton distance (1.4 Bohr) \cite{delaney06}, necessary for the system being metallic. 
For $r_s=1.50$ and 1.60 we also fit $\sigma(\omega)$ points to
the Drude formula. The fitted curves become flatter as the
density decreases, which signals that at $r_s=1.60$, liquid
hydrogen has a small electronic relaxation time, making it a
bad metal, similar to liquid carbon \cite{galli90}.

Finally, we explicitly test the importance of finite-size
effects 
by performing simulations for 54-atom cells as shown in
Fig. \ref{sigmaandDOSat3KNP54}.
The energy gap at $r_s=1.65$ that is present in the 32 atom
simulation now disappears. The liquid semiconductor [at
$r_s=1.75$ the extrapolated DC conductivity is around 1300
$(\Omega$ cm$)^{-1}$ less than a typical liquid metal value
\cite{shimoji77, mott71}, hence the name] to liquid metal
transition density is estimated to be in the range $1.65 \leq
r_s \leq 1.75 $, which is again close to the experimental
density of $r_s=1.62$ \cite{weir96}. Further increase of system
size is not possible at present, but previous CEIMC
calculations have found that the ground-state energy with 54
and 108 atoms are very close if twist-averaged boundary
conditions are used. Therefore, we expect that the liquid metal
transition density determined above is close to the
thermodynamic limit. A more definitive answer will require
significant additional calculations.

We have proposed a method to calculate the frequency dependent
electrical conductivity using CEIMC with correlated Monte Carlo
sampling and the many-body Kubo formula. The method is able to
estimate some of the lowest-lying energies and their
corresponding overlap matrices, and hence it is suitable for
the extrapolation of DC conductivity from the AC conductivity.
With this method we study the DC conductivity of liquid
hydrogen as a function of density at 3000 K and at high
pressure, and show the metallization. Our DC conductivity
values at the metallization point is in good agreement with the
shock-wave experiments \cite{weir96}. Furthermore, our
metallization density ($r_s\sim 1.65$) is close to the
experimental value of $r_s=1.62$. In the future, it will be
interesting to apply the method to more densities and
temperatures to have a complete understanding of liquid
hydrogen electronic transport properties. It is also very
promising to extend our method to other systems, such as
lattice models.

F.L. thanks Michele Casula for helpful
discussions. C.P. thanks the Institute of Condensed
Matter Theory at the University of Illinois at Urbana-Champaign
for a short term visit. This research was sponsored in part by the
National Nuclear Security Administration under the Stewardship Science
Academic Alliances program through DOE Grant DE-FG52-06NA26170;
Ministero dell'Istruzione, dell'Universita e della Ricerca Grant
PRIN2007 (to C.P.). Computer time was provided by NCSA  at the University of
Illinois at Urbana-Champaign and CINECA, Italy.


\begin{thebibliography}{99}
\bibitem{weir96}S. T. Weir, A. C. Mitchell, and W. J. Nellis, \prl {\bf 76}, 1860 (1996);
W. J. Nellis, S. T. Weir, and A. C. Mitchell, \prb {\bf 59}, 3434 (1999).

\bibitem{hohl97}O. Pfaffenzeller and D. Hohl, J. Phys.: Condens. Matter {\bf 9}, 11023 (1997).

\bibitem{collins01}L. A. Collins \etal, \prb {\bf 63}, 184110 (2001).

\bibitem{redmer08}B. Holst, R. Redmer, and M. P. Desjarlais, \prb {\bf 77}, 184201 (2008).

\bibitem{pierleoni04}C. Pierleoni, D. M. Ceperley, and M. Holzmann, \prl {\bf 93}, 146402 (2004);
C. Pierleoni and D. M. Ceperley, Chem. Phys. Chem. {\bf 6}, 1872 (2005); {\it ibid},
{\it Lecture Notes in Physics} {\bf 703}, 641 (2006).

\bibitem{ceperley88}D. M. Ceperley and B. Bernu, J. Chem. Phys. {\bf 89}, 6316 (1988);
                    B. Bernu, D. M. Ceperley, and W. A. Lester Jr., J. Chem. Phys. {\bf 93}, 552 (1990).

\bibitem{hammond94} B. L. Hammond, W. A. Lester Jr., and P. J. Reynolds, {\it Monte Carlo methods in ab
initio quantum chemistry}, Chap. 6, World Scientific, Singapore (1994).

\bibitem{delaney06}K. T. Delaney, C. Pierleoni, and D. M. Ceperley, \prl {\bf 97}, 235702 (2006).

\bibitem{baroni99}S. Baroni and S. Moroni, \prl {\bf 82}, 4745 (1999).

\bibitem{kubo57}R. Kubo, J. Phys. Soc. Japan {\bf 12}, 570 (1957).

\bibitem{lin01}C. Lin, F. H. Zong, and D. M. Ceperley, \pre {\bf 64}, 016702 (2001).

\bibitem{galli90}G. Galli, R. M. Martin, R. Car, and M. Parrinello, \prb {\bf 42}, 7470 (1990).

\bibitem{pierleoni08}C. Pierleoni, K. T. Delaney, M. A. Morales, D. M. Ceperley, and M. Holzmann,
Comp. Phys. Comm. {\bf 179}, 89 (2008).

\bibitem{galli00}G. Galli, R. Q. Hood, A. U. Hazi, and F. Gygi, \prb {\bf 61}, 909 (2000).

\bibitem{shimoji77}M. Shimoji, {\it Liquid Metals}, p. 381, Academic Press, London (1977).

\bibitem{mott71}N. F. Mott, Philos. Mag. {\bf 24}, 2 (1971).

\end{thebibliography}
\end{document}